\documentclass[twocolumn,showpacs,amsmath,amssymb]{revtex4}

\usepackage{graphics}
\usepackage[dvips]{graphicx}

\def\sqr#1#2{{\vcenter{\hrule height.#2pt\hbox{\vrule width.#2pt height#1pt
\kern#1pt \vrule width.#2pt}\hrule height.#2pt}}}
\def\square{\mathchoice\sqr64\sqr64\sqr{4.2}3\sqr{3.0}3}

\begin{document}

\title{Newtonian limit of scalar-tensor theories and galactic dynamics: isolated and  
interacting galaxies}

\author{Jorge ~L. ~Cervantes--Cota$^*$, M. A.~ Rodr\'\i guez--Meza, R.  Gabbasov, 
 and J. Klapp} 
\address{Depto. de F\'{\i}sica, Instituto Nacional de Investigaciones
Nucleares, Apo.d Postal 18-1027, M\'{e}xico D.F. 11801, M\'{e}xico. \\
$^*$ e-mail: jorge@nuclear.inin.mx}

\date{\today}

\begin{abstract}
We use  the Newtonian limit of  a general scalar-tensor theory  around a background 
field to study astrophysical effects. The  gravitational  theory modifies the standard 
Newtonian potential  by adding a Yukawa term to it, which is  quantified by two 
theoretical parameters: $\lambda$, the  lenghtscale of the  gravitational interaction and 
its strength, $\alpha$.  Within this formalism  we firstly present  a numerical study on the 
formation of bars  in isolated galaxies.  We have found for positive $\alpha$ that
the modified gravity destabilizes the galactic discs and leads to rapid bar formation in
isolated galaxies. Values of $\lambda$ in the range
$\approx 8$ -- 14 kpc produce strongest bars in isolated models.  Then, we extent this work by 
considering tidal effects due to interacting galaxies.  We send two spirals to collide and  study the 
bar properties of the remnant. We  characterize the bar kinematical properties in terms of our 
parameters ($\lambda, \alpha$).    


\bigskip

Usamos el l\'{i}mite newtoniano de una teor\'{i}a  escalar-tensorial general alrededor de un campo del 
fondo para estudiar  efectos astrof\'{i}sicos. La teor\'{i}a gravitacional  modifica el potencial newtoniano 
est\'andar, agregandole un t\'ermino de Yukawa, el cual se cuantifica por dos par\'ametros 
te\'oricos: $\lambda$, la escala de longitud de la interacci\'on gravitacional y  $\alpha$, su intensidad.   
 Dentro  de este formalismo primero presentamos  un estudio num\'erico 
de la formaci\'on de barras en galaxias aisladas.  Encontramos que para $\alpha$
positiva la gravedad modificada desestabiliza a los discos gal\'acticos y lleva a una
r\'apida formaci\'on de la barra en galaxias aisladas. Valores de $\lambda$ en el
rango de $\approx 8$ -- 14 kpc producen barras m\'as pronunciadas en los modelos
de galaxias aisladas. Despu\'es, extendemos este trabajo al considerar 
efectos de marea debido a galaxias que interaccionan. Enviamos dos galaxias espirales a colisionar  para estudiar 
las caracter\'{i}sticas de la barra del remanente formado. Caracterizamos las propiedades cinem\'aticas de la barra en 
t\'erminos de nuestros par\'ametros ($\lambda, \alpha$).    

\medskip
\noindent {Descriptores}:  Formaci\'on de la barra; interacci\'on de galaxias;
teor\'{i}a  escalar-tensorial 
\end{abstract}


\maketitle

\section{Introduction}
In recent years there have been some attempts to explain observed gravitational 
effects that imply the existence of dark matter (DM) and dark energy (DE).  These effects go from 
galactic to cosmological scales and we still do not know the nature of DM and DE, or if they 
are related to each other, or even if DM is unique or there is a set  of dark 
multi-components.  One possibility is that scalar fields (SF) play a role in the modified  
dynamics.   Particularly, SF are often applied in cosmology to accomplish 
the Universe's accelerated expansion, inferred from  Supernovae Ia redshifts, CMB Doppler 
peaks measurements, large-scale structure surveys and cosmological simulations  
\cite{Ri98,Pe99,deB00,Ri01,Pe02,Ef02,Be03,Sp06}. For a review of  all these topics see for  
instance \cite{BrCeSa04}.   

The way in which SF couple to gravity is also unknown, simply because there is a lack of a 
unique fundamental  theory that  explains the intricate relationship of matter and  its gravitational 
background.   One possibility is that SF are coupled non-minimally to gravity, at a Lagrangian 
footing, as it happens when string theories are compactified to four space-time dimensions 
\citep{Gr88}. The resulting effective theory is a  scalar-tensor theory (STT) of gravitation,  that 
can generically be described by  arbitrary scalar functions, apart from the 
geometrical part \cite{BrDi61,Wa70}. In the  past we have studied different effects of this 
type of theories  in cosmology \cite{ChCeNu92,CeDe05a,CeDe05b,ChCe95,CeCh99,Ce99}, and 
more recently  we have considered the Newtonian limit of STT and apply it to 
astrophysical phenomena.  We  have
computed potential-density pairs for various halo density
profiles \cite{RoCe04} and axisymmetric systems  \cite{Ro05}. It was found that the augment of rotation curves and
velocity dispersions depend on the parameters of the SF. On the
other hand, in Ref.  \cite{Rodriguez-Mezaothers01} we have computed the
effect of SF on the transfer of angular momentum between protogalaxies.  In the present work,  we pursue to 
study bars in spiral galaxies.   Observations 
of spiral galaxies indicate that the presence of a
 bar is a common feature \cite{ElmegreenElm83}. Instabilities in
 isolated stellar and gaseous discs lead to bar formation; see \citep{Toomre64} for pioneer studies and 
 \citep{Ba98,Gabbasov2006} for a modern view. The bar formation in isolated models
 has been widely studied both analytically and numerically  \citep[e.g.,][]{Hohl71,Sellwood81, SellwoodCarl84,
SellwoodAthan86,AthanSell86,Weinberg85,DebattistaSell00,WeinbergKatz02} and it is  studied 
using the above STT formalism in the first part of this paper. In the second part,  we consider in dynamical effects of 
non-isolated systems which are found in 
clusters of galaxies. In this sense, it has been suggested that the observed bar in many
spirals is the result of the gravitational interaction between two
or more nearby galaxies. For instance, \citet{Nogushi87} has found
that during the collision of two galaxies and between the first
and the second closest approaches, the disc takes a transient bar
shape. The gravitational interaction between the two galaxies
gives rise to perturbations in the orbits of the stars that results in the formation of the bar.

 Bar formation in stellar discs depends upon various simultaneous effects. In the case of 
 collisions, the 2D-simulations have shown that these factors are \citep{Salo90}: rotation 
 curve shape, disc-halo mass ratio, perturbation force and geometry. Additionally, simulations 
 suffer from numerical effects such as low
 spatial and temporal resolution, too few particles representing
 the system, and an approximate force model. These effects were
 studied by us in Refs. \citep{Gabbasov2006a,Gabbasov2006},
 where  it was shown that  specific parameter choices  may change bar 
 properties. Once numerical effects are controlled, we may investigate all 
 the other model parameters, which in our case are ($\lambda$, $\alpha$).

In the present paper we study the formation of bars as a product of both 
instabilities of isolated galaxies and as a result of the collision of two spirals in 
framework of  STT. In particular, we consider a non-minimally coupled SF in the
Newtonian limit  (section \ref{STT_section})  and use  the resulting  modified gravitation
force in our 3D-simulations. In this way, 
all collisionless particles mutually interact with the modified 
 gravitational force. Then, we investigate isolated galaxies (section \ref{isolated_galaxies}) 
 and head-on and off-axis impacts  of two disc galaxies and the properties of tidally formed bars
 (section \ref{interact_galaxies}).  We finally draw some conclusions (section \ref{conclusions}).
 
\section{Scalar--tensor theory and its Newtonian limit}\label{STT_section}

A typical STT is given by the following Lagrangian \cite{BrDi61,Wa70}:
\begin{equation}
\label{SFLagrangian}
    {\cal L} = \frac{\sqrt{-g}}{16\pi} \left[
    -\phi R + \frac{\omega(\phi)}{\phi} (\partial \phi)^2 - V(\phi)
    \right] + {\cal L}_M(g_{\mu\nu}) \; ,
\end{equation}
from which we get the gravity and SF equations.
 Here $g_{\mu\nu}$ is the metric, ${\cal L}_M(g_{\mu\nu})$ is the
 matter Lagrangian and $\omega(\phi)$ and
 $V(\phi)$ are arbitrary functions of the SF. Thus the
 gravitational equation is
\begin{eqnarray}
\label{Einsteineqn}
    R_{\mu\nu}-\frac{1}{2}g_{\mu\nu}R&=&\frac{1}{\phi}\left[8\pi
    T_{\mu\nu}+\frac{1}{2}V
    g_{\mu\nu}+\frac{\omega}{\phi}\partial_{\mu} \phi\ \partial_{\nu}
    \phi
    \right. \\ & & \left.
    -\frac{1}{2}\frac{\omega}{\phi}(\partial_{\mu} \phi)^2
    g_{\mu\nu}+{\phi_{;\mu\nu}}-g_{\mu\nu}\square\,\phi
\right].   \nonumber
\end{eqnarray}
The SF part is described by the following equation
\begin{equation}
\label{SFparteqn}
    \square\,\phi + \frac{\phi V'-2V}{3+2\omega}=
    \frac{1}{3+2\omega}\left[8\pi T -{\omega}' (\partial \phi)^2\right]  \, ,  
\end{equation}
where a prime (') denotes the derivative with respect to SF  ($\phi$).

 In accordance with the Newtonian approximation, gravity and SF are
 weak. Then, we expect to have small deviations of the SF 
 around the background field.  Assuming also that the velocities of stars and DM particles 
 are non-relativistic, we perform the expansion of the field equations around the
background quantities $\langle\phi\rangle$ and $\eta_{\mu\nu}$.  Even though  
the expansion of the above equations to first order is well known  \citep{No70,Helbig91,Wi93}, we 
explicitly show it  in the appendix since our definition of the background field is 
$\langle \phi \rangle = G_{N}^{-1}(1+\alpha)$, which is non-trivial, 
and this changes some constant terms in the equations.  Accordingly, we 
obtain Eqs. (\ref{fieldeq1-limit}) and (\ref{fieldeq2-limit}): 
\begin{eqnarray}
\frac{1}{2} \nabla^2 h_{00} &=& \frac{G{_N}}{1+\alpha} \left[ 4\pi
\rho - \frac{1}{2} \nabla^2 \bar{\phi}\right]  \; ,
\label{pares_eq_h00}\\
   \nabla^2 \bar{\phi} - m^2 \bar{\phi} &=&
- 8\pi \alpha \rho \; , \label{pares_eq_phibar}  
\end{eqnarray}
where $\rho$ is matter density of  DM or stars stemming from the  
energy-momentum tensor,  
$G_N$ is the Newtonian gravitational constant and  
$\alpha \equiv 1/(3 + 2 \omega)$ is a constant,  in which    
$\omega$ is the Brans--Dicke parameter \citep{BrDi61},
here defined in theories that include scalar potentials. 
Equations (\ref{pares_eq_h00}) and (\ref{pares_eq_phibar})
represent the Newtonian limit of a set of STT with arbitrary potentials ($V(\phi)$) 
 and functions $\omega(\phi)$ that are Taylor expanded around some 
value.  The resulting equations are thus  
distinguished by the constants $\alpha$ and $m$. 

In the above expansion we have set the cosmological constant equal
to zero since within galactic scales its influence is negligible.
This is because the average density in a galaxy is much larger
than a cosmological constant that is compatible with observations.
Thus, we only consider the influence of luminous and dark matter.
These matter components gravitate in accordance  with the
modified--Newtonian theory determined by Eqs.~(\ref{pares_eq_h00})
and \ (\ref{pares_eq_phibar}). The latter is a Klein-Gordon equation 
which contains an effective mass $m$ term, whose Compton
wavelength ($\lambda = h/m c$) implies a length scale for the modified 
dynamics. We shall assume this scale to be of the order of tens of
kilo-parsecs, which corresponds to a very small mass, $m\sim
10^{-26}$ eV.

Note that Eq.~(\ref{pares_eq_h00}) can be cast as a Poisson
equation for $\psi \equiv (1/2) ( h_{00} + \bar{\phi}/\langle \phi
\rangle)$,
\begin{equation}
\nabla^2 \psi =  4\pi G{_N} \,  \rho /(1+\alpha) \; , \label{pares_eq_psi}
\end{equation}
Thus, the modified Newtonian potential is now
given by
\begin{equation} \label{phi-new}
\Phi_N \equiv \frac{1}{2} h_{00} = \psi - \frac{1}{2}
\frac{\bar{\phi}}{\langle \phi \rangle} \, .
\end{equation}

Particular solutions, the so-called potential--density pairs
\citep{BT94}, were recently found for the NFW's and Dehnen's
density profiles \citep{RoCe04} and for axisymmetric
systems \citep{Ro05}. For point masses (of non-SF nature) the solution is
well known \citep{Helbig91,FischbachT98} and here is adapted to our definition
of the background field, $\langle \phi \rangle = G_{N}^{-1}(1+\alpha)$:
\begin{eqnarray}
\bar{\phi} = 2 \alpha u_\lambda \, , \quad
\Phi_N = -u - \alpha u_\lambda \,  ,
\end{eqnarray}
where
\begin{eqnarray}
u &=& \frac{G_N}{(1+\alpha)} \sum_s \frac{m_s}{| {\bf r} - {\bf r}_s |} \, , \\
u_\lambda &=& \frac{G_N}{(1+\alpha)}   \sum_s \frac{ m_s}{| {\bf r} - {\bf r}_s |} {\rm
e}^{ -| {\bf r} - {\bf r}_s |/\lambda } \; ,
\end{eqnarray}
with $m_s$ being a source mass.   
The potential $u$ is the Newtonian part and $u_\lambda$ is the SF
modification  which is of Yukawa type.  The total 
gravitational force on a particle of mass $m_i$ is
\begin{equation}
{\bf F} = -m_i \nabla \Phi_N = m_i {\bf a} .
\end{equation}
Thus, gravitating particles, that in our simulations are stars or DM particles,  
will feel the influence of Newtonian gravity ($u$) plus a SF force due to the term 
$u_\lambda$. 

The gravitational potential of a single particle arising from 
the above formalism is: 
\begin{equation} \label{phi_New}
\Phi_N = \frac{G_N m_{s}}{(1+\alpha) \, r} (1+ \alpha e^{-r/\lambda}) \, ,
\end{equation}
For local scales, $r \ll \lambda$, deviations from the Newtonian
theory are exponentially suppressed, and for $r \gg \lambda$ the
Newtonian constant diminishes (augments) to $G_{N}/(1+\alpha)$ for
positive (negative) $\alpha$.   This means that equation (\ref{phi_New}) 
fulfills all local tests of the Newtonian dynamics, and it is only constrained 
by experiments or tests on scales larger than -or of the order of- $\lambda$, which in our case 
is of the order of galactic scales.  By contrast, if one defines $<\phi> \equiv 1/G_N$, then
the effective Newtonian constant is modified at scales $r < \lambda$, and  stringent, local 
constraints applies, demanding  $\alpha$ to be less than $10^{-10}$ \cite{FischbachT98}. 
 This latter approach will not be considered here.  

Recently, the effect of STT has been investigated in different
cosmological scenarios in which variations of the Newtonian
constant are constrained from a phenomenological point of view.  
For instance, \cite{UmIcYa05} studied
the influence of varying $G_N$ on the Doppler peaks of the CMBR,
and concluded that their parameter ($\xi= G/G_N$) can be in the
interval $0.75 \le \xi \le 1.74$  to be within the error bars of
the CMBR measurements. In our notation this translates into $-0.43
\le \alpha \le 0.33$.  However, this range for $\alpha$ has to be
taken as a rough estimation, since these authors have only
considered  a variation of $G_N$, and not a full perturbation
study within STT. The latter has been done by \cite{NaChSu02}, who
found some allowed deviations from the Newtonian dynamics, that
translated into our parameter is $\alpha = 0.04$; however, a
comparison with observations in not made. On the other hand, a
structure formation analysis has been done in \cite{ShShYoSu05},
in which deviations of the matter power spectrum are studied by
adding a Yukawa potential to the Newtonian. They found some
allowed dynamics, that turns out to constrain our parameter to be
within $-1.0 \le \alpha \le 0.5$; but again a self-consistent 
perturbation study in general STT is missing. Thus, the above
three estimates can be taken as order-of-magnitude constraints for
our models.  Note that even when it is not theoretically justified to take 
negative values  for  $\alpha$, phenomenology admits them.  
In this work, however, we only consider positive values of $\alpha$.

%
\section{isolated galaxy simulations}\label{isolated_galaxies}
We use the standard procedure to
construct a galaxy model with a Newtonian potential described in
\citep{Gabbasov2006a, Gabbasov2006}. The galaxy consists of a disc,
halo, and bulge and its  initial condition
is constructed using the Hernquist halo model (a Dehnen's family member with
$\gamma=1$, see \citep{RoCe04}).  To perform the 3D-simulations we used the 
{\it gbsph} code  ({\texttt{www.astro.inin.mx/mar/nagbody})   modified to include the contribution
 of the scalar fields as given in the preceding section. 
 The forces were computed with a tolerance parameter $\theta=0.75$,
 and including the quadruple term. We use Barnes's 
model parameters  and system of units \cite{Ba98}.  The mass, length and time scales are set to 
$2.2 \times 10^{11} \,  {\mbox M}_{\odot} = 1.40 \,  {\rm kpc} = 1$ 
and 250 Myr$ =1$, respectively. In these units, the gravitational constant is $G_N = 1$. 
The discÕs scale height is $z_0 = 0.007$ and the half mass radius 
of the galaxy is located at  $R_{1/2} \approx 11$ kpc.

All isolated runs were performed with $\varepsilon=0.015$ ($=0.6$ kpc) and
$\Delta t=1/128$ for $N=40\,960$, and $\varepsilon=0.008$ and
$\Delta t=1/128$ for $N=163\,840$, respectively. Galaxies were
evolved up to $t=12$ ($3$ Gyrs). Results of some of the runs are
summarized in Table \ref{SFgalaxy_test}, where
columns are:
the model label
 (1), the number of particles (2), the
 SF strength $\alpha$ (3), and SF length scale $\lambda$ (4). As a result of
 simulations the following control parameters are displayed:
 the relative change of components of the disc velocity dispersions, measured
 at time 0.5 and 3 Gyrs  (5-7) \citep{Gabbasov2006a},
 the disc angular momentum loss (8), the Toomre's Q parameter (9), the Toomre's $X$
 parameter (10). The expressions for the last two parameters can be found in
 Ref.\ \citep{BT94}.

Table \ref{SFgalaxy_test} shows, as in previous results \citep{Gabbasov2006a},
that experiments made with $N=163\,840$ is less collisional than with
smaller number of particles (compare columns 5--7). Runs series SFB are computed
with SF strength $\alpha=0.1$ and series SFC with $\alpha=0.3$ and we observed
that the heating of the disc is also higher for a higher SF strength.

In Fig.\ 1 we show the time evolution
of the amplitude of the second harmonic, $|A_2|$, which tells us about the appearance of
the bar at approximately 1.5 Gyr. The bar is stronger for $\alpha=0.3$ than $\alpha=0.1$, meanwhile
in run SFB00 (Newtonian) a bar appears only at $t\approx 4$ Gyrs. Also,
the disc in presence of SF
heats stronger than in Newtonian case. This is due to a bar that
appears in all simulations with SF.

%
\begin{table}
 \caption{Numerical parameters of galaxy evolution runs.}
\label{SFgalaxy_test}
 \centering
  \begin{tabular}{@{}lccccccccc@{}}
  \hline\hline
  (1) & (2) & (3) & (4) & (5) & (6) & (7) & (8) & (9) & (10)  \\
  \hline
Model & $N$ & $\alpha$ & $\lambda$ & $\gamma_r$ &
$\gamma_{\varphi}$ & $\gamma_{z}$ & $\frac{\Delta
L_{\mathrm{d}}}{L_{\mathrm{d 0}}}\times 100$ & $Q$ & $X_2$  \\
\hline
 SFA00 & 40\,960 & 0.0 & - & 0.724 & 0.734 & 1.037 & 4.1 & 2.5 & 2.2  \\
 SFA01 & " & 0.3 & 1.0 & 1.096 & 0.933 & 1.404 & 6.1 & 3.0 & 2.0  \\
 SFA02 & " & 0.3 & 0.4 & 1.051 & 0.936 & 1.262 & 4.5 & 2.8 & 2.0  \\
 SFA03 & " & 0.3 & 0.2 & 2.049 & 1.576 & 1.608 & 8.4 & 3.0 & 1.5  \\
 SFA04 & " & 0.3 & 0.1 & 1.673 & 1.345 & 1.114 & 7.7 & 2.9 & 1.0  \\
\hline
 SFB00 & 163\,840 & 0.0 & - & 0.579 & 0.447 & 0.494 & 1.1 & 2.0 & 3.0  \\
 SFB01 & " & 0.1 & 1.0 & 0.898 & 0.687 & 0.636 & 2.7 & 2.5 & 3.0   \\
 SFB02 & " & 0.1 & 0.4 & 1.176 & 0.798 & 0.662 & 3.3 & 2.4 & 2.6   \\
 SFB03 & " & 0.1 & 0.2 & 1.058 & 0.805 & 0.617 & 2.9 & 2.3 & 2.5   \\
 SFB04 & " & 0.1 & 0.1 & 0.916 & 0.749 & 0.624 & 2.3 & 2.2 & 2.7   \\
 \hline
 SFC01 & " & 0.3 & 1.0 & 0.889 & 0.653 & 0.593 & 4.4 & 2.5 & 2.5   \\
 SFC02 & " & 0.3 & 0.4 & 1.023 & 0.785 & 0.572 & 5.6 & 2.6 & 2.5   \\
 SFC03 & " & 0.3 & 0.2 & 1.779 & 1.325 & 0.764 & 8.8 & 3.0 & 1.7   \\
 SFC04 & " & 0.3 & 0.1 & 1.279 & 1.015 & 0.613 & 7.7 & 2.6 & 1.2   \\
  \hline
\end{tabular}
\end{table}
\begin{figure}
\begin{center}
\includegraphics[width=75mm]{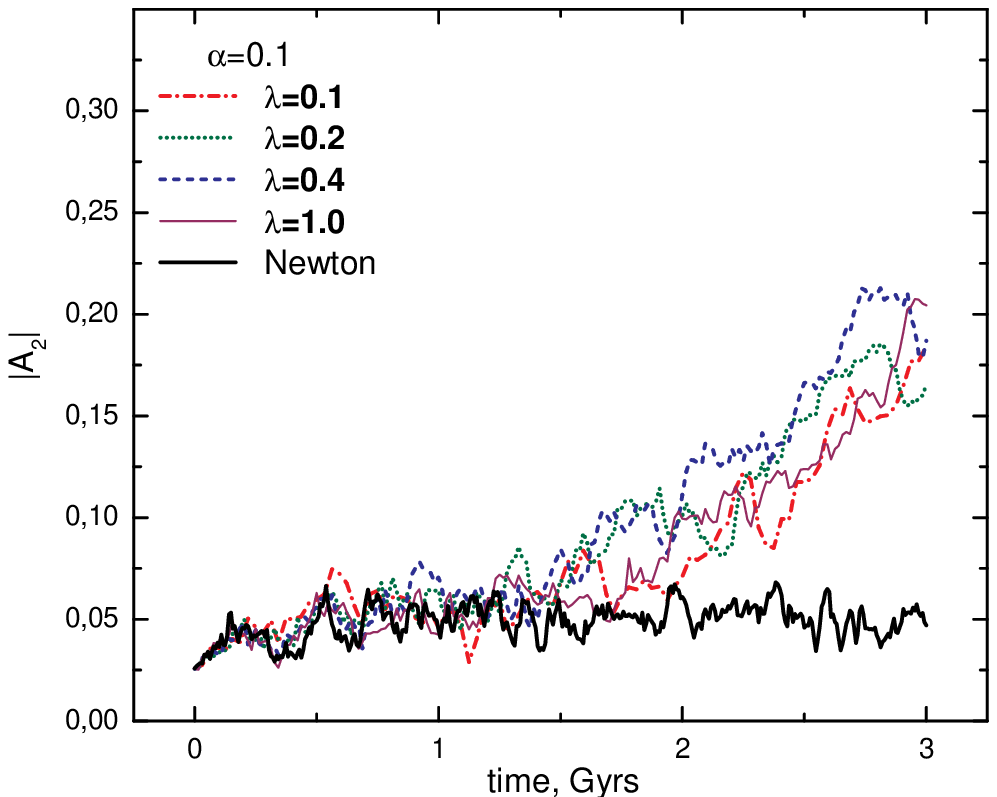}
\includegraphics[width=75mm]{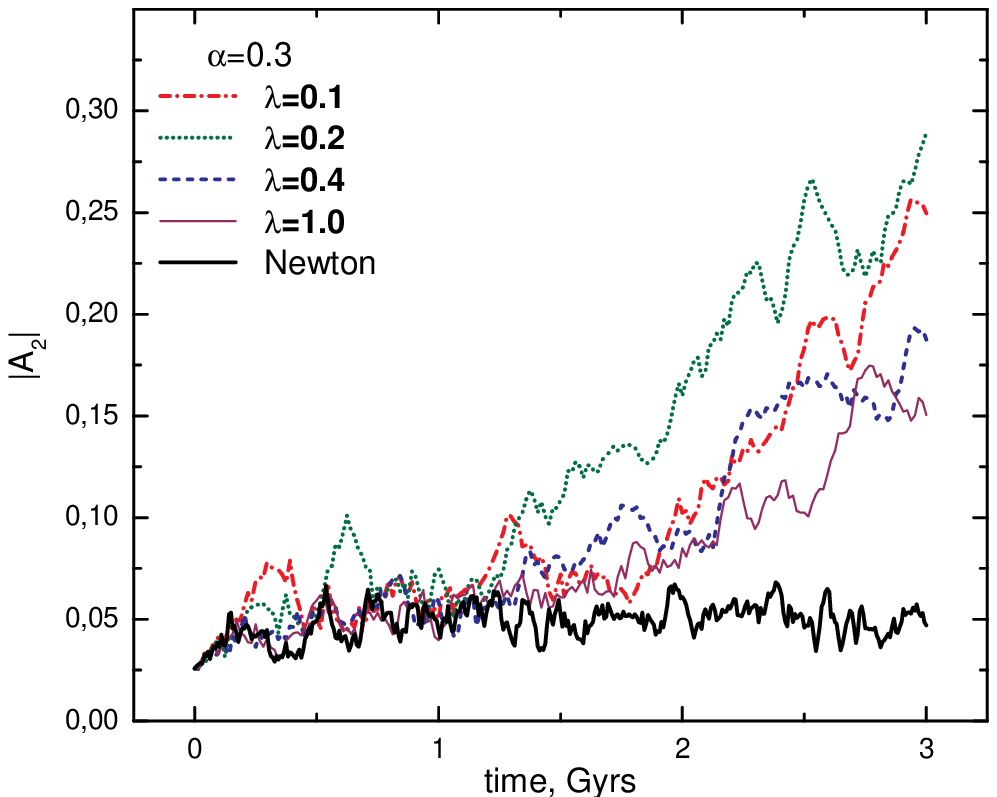}
\caption{Evolution of $|A_2|$ for models SFB00-SFB04 (top panel)
and for models SFC01-SFC04 (bottom panel).} \label{sf-isol-harms}
\end{center}
\end{figure}
%
\begin{figure}[b]
\begin{center}
\includegraphics[width=75mm]{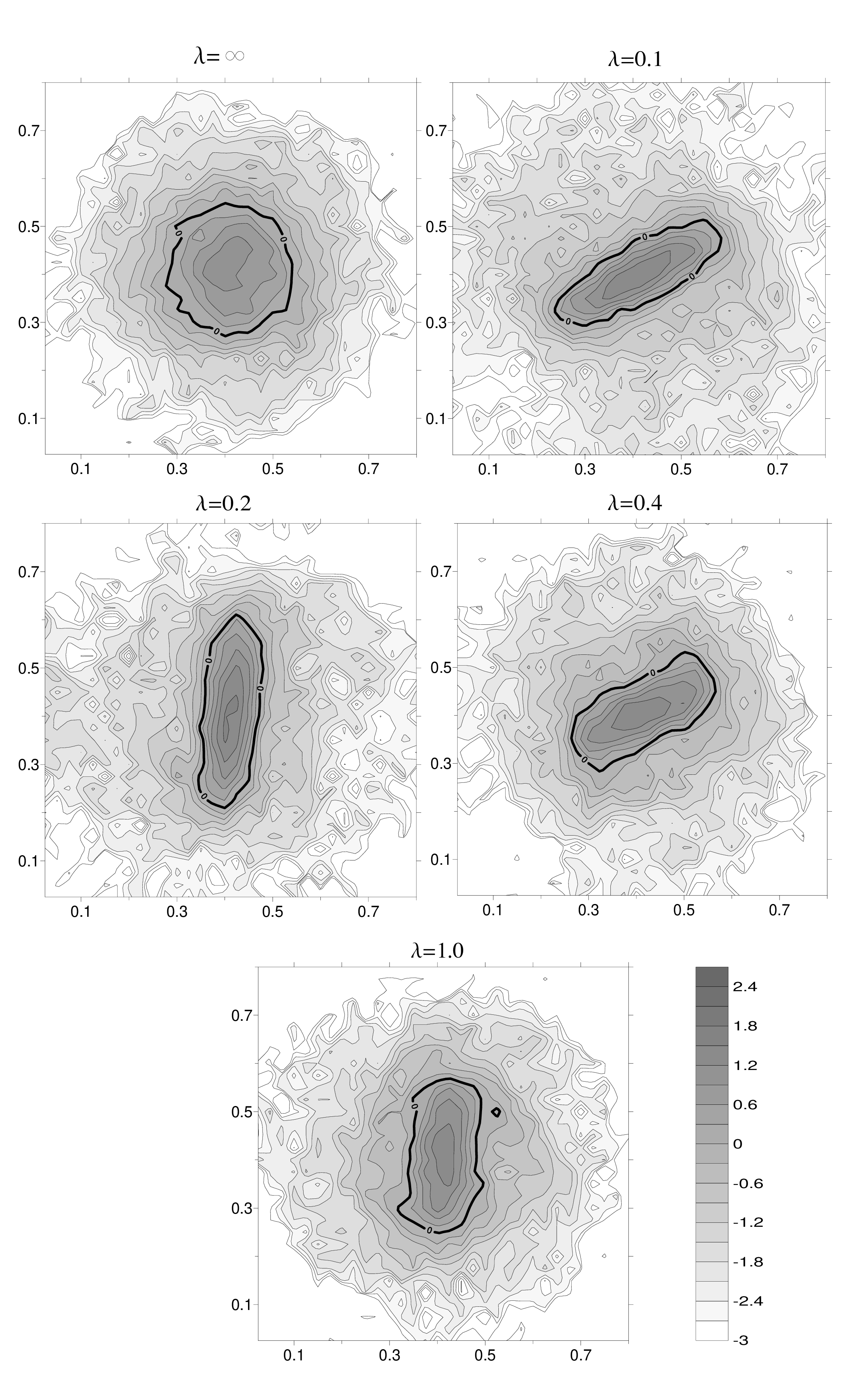}
\caption{Projected disc density contours at $t=3$Gyrs for models SFB00, SFC01-SFC04. Thick lines indicate 
zero level density contours in logarithmic scale.} \label{fig-contours}
\end{center}
\end{figure}
%

As it can be seen, a stronger SF produce stronger bars for
intermediate $\lambda$, reaching $|A_2|\approx 0.3$ for models
SFC02 and SFC03. We think that the enhanced heating and transfer
of the disc angular momentum in runs SFB02 and SFC03 is due to a
stronger and larger bar which in turn depends on SF parameters.
The bar angular velocities neither depend on $\lambda$ nor
$\alpha$ and have initially the value $\Omega_p\approx 6$, which
after $\sim 1$ Gyrs decrease to $\Omega_p\approx 5.7$ in code
units.

From Fig.~\ref{sf-isol-harms} it is seen that SF with $\alpha=0.3$
and $\lambda=0.2$ produces the strongest bar.  This is probably due
to some resonance of disc particle orbits with the selected
$\lambda$, which is roughly equal to the bar's length. The same result can be seen in 
Fig.~\ref{fig-contours}, where the final density contours of discs are plotted.  

\section{Interacting galaxies simulations}\label{interact_galaxies}
In this section we describe 3D-simulations of collisions of two identical galaxies, each of which are the 
same as the used in study of isolated galaxies in the preceding section.  
In a recent paper \cite{Gabbasov2006a} we have shown how the numerical parameters  influence the bar properties.
In accordance with this study we choose for the total number of particles $N = 163\,840$, 
the softening parameter $\varepsilon=0.008$ and the time-step $\Delta t=1/128$. This choice of 
parameters prevents from an early bar formation, hence, it permits to study the tidal effects on the bar formation.

We study the effect of the SF on tidal bar formation and its properties, such as its amplitude and
rotational velocity. 
We compare the bar amplitude and its pattern speed for fixed $\alpha$, varying $\lambda$ and the
 impact parameter ($p$).

In order to maintain the same impact velocity and pericentric
separation we have studied head-on and off-axis impacts of
galaxies launched with the initial velocities $v=|v_x|= 200$ km/s 
and the impact parameter $p$, whose values are listed
in Table~\ref{SF-collisions}. The galaxies were relaxed up to
$t=0.25$ Gyrs before placed on the orbits with the initial separation
$R=64$ kpc. The whole collision is followed up to $t=4$ Gyrs. We
consider prograde-retrograde and planar collisions which allow us
to investigate two possible directions of rotation and to check
whether the bars emerge in retrograde discs during the violent
collision. The first galaxy is retrograde, moves to the left and
for off-axis collisions is placed above, whereas the second galaxy
is prograde, moves to the right and is located below the first
one.

The performed collision simulations are summarized in
Table~\ref{SF-collisions}, where we have varied $\lambda$ for a fixed value
of the Yukawa strength  $\alpha = 0.1$.  We use the following labeling in
model names: SF - for the scalar field, L(xx) - for the lambda
value multiplied by ten, and P(y) - for the pericentric parameter
expressed in number of disc radii. For the Newtonian simulations the 
SFN label is used. In all runs the total energy and the total
angular momentum conserve better than $1$ \%. Movies of some
collision simulations are available at {\texttt{www.astro.inin.mx/ruslan/stt.}}

\begin{table}
  \caption{Parameters of collisions with fixed  $\alpha = 0.1$}
  \label{SF-collisions}
\smallskip
  \centering
  \begin{tabular}{@{}lcccc@{}}
  \hline \hline
Run & $\lambda$ &   $ p $   & Wiggle ? & Wiggle ? \\
  &  &  & Disc 1 & Disc 2 \\
\hline
 SFL01P0 & $\quad 0.1\quad $ & $\quad 0\quad $ & no & no \\
 SFL01P1 & - & 0.4 & yes & no \\
 SFL01P2 & - & 0.8 & yes & yes \\
\hline
 SFL02P0 & 0.2 & 0 & no & no \\
 SFL02P1 & - & 0.4 & yes & no \\
 SFL02P2 & - & 0.8 & yes & yes \\
\hline
 SFL04P0 & 0.4 & 0 & no & no \\
 SFL04P1 & - & 0.4 & yes & no \\
 SFL04P2 & - & 0.8 & yes & yes \\
\hline
 SFL10P0 & 1.0 & 0 & no & no \\
 SFL10P1 & - & 0.4 & yes & no \\
 SFL10P2 & - & 0.8 & no & yes \\
\hline
 SFNP0 & $\infty$ & 0 & no & no \\
 SFNP1 & - & 0.4 & yes & yes \\
 SFNP2 & - & 0.8 & no & yes \\
 \hline
\end{tabular}
\end{table}

For all numerical experiments, 
we have plotted  the evolution of the amplitude 
of the second harmonic, $|A_2|$, which indicates the presence of a 
bar and corresponding pattern velocity, $\Omega$.
  We first consider head-on collisions. The graphics of $|A_2|$ and $\Omega$ shown in
Figs.~\ref{sf-harms0R} and \ref{sf-omegas0R}, respectively, are
all similar and comparable to the Newtonian case, except for small
oscillations at the end of the run. These oscillations increase
with increasing $\lambda$ and are also present in plots of bar
pattern speed. 

Next, we discuss si\-mu\-lations with an impact parameter equal to the disc's radius, 
$p=16$ kpc. A striking difference between
the run SFNP1 and runs SFL01P1-SFL10P1 is that for SF models the bars in both discs
have roughly the same amplitude, independently of $\lambda$, whereas in Newtonian case 
their amplitudes differ by roughly twice, see  Fig.~\ref{sf-harms1R} .
The fact that the retrograde discs form bars indicate that the discs in presence of SF are unstable and a 
short and strong enough perturbation is sufficient to produce a bar.
As in Newtonian case, the retrograde bars are slightly faster than the prograde
ones, indicating their similarity with isolated bars. In general, the pattern velocities of the prograde 
and retrograde bars for models with SF are smaller than for the Newtonian model, see Fig.\ref{sf-omegas1R} .

Concerning encounters with $p=32$ kpc, the curves in Figs.~\ref{sf-harms2R} and
\ref{sf-omegas2R} are similar for each case and show no much difference. The only remarkable
feature is the higher peaks in amplitudes of the prograde discs with SF in comparison with the Newtonian
case.

%
\begin{figure}[t]
\begin{center}
\includegraphics[width=65mm]{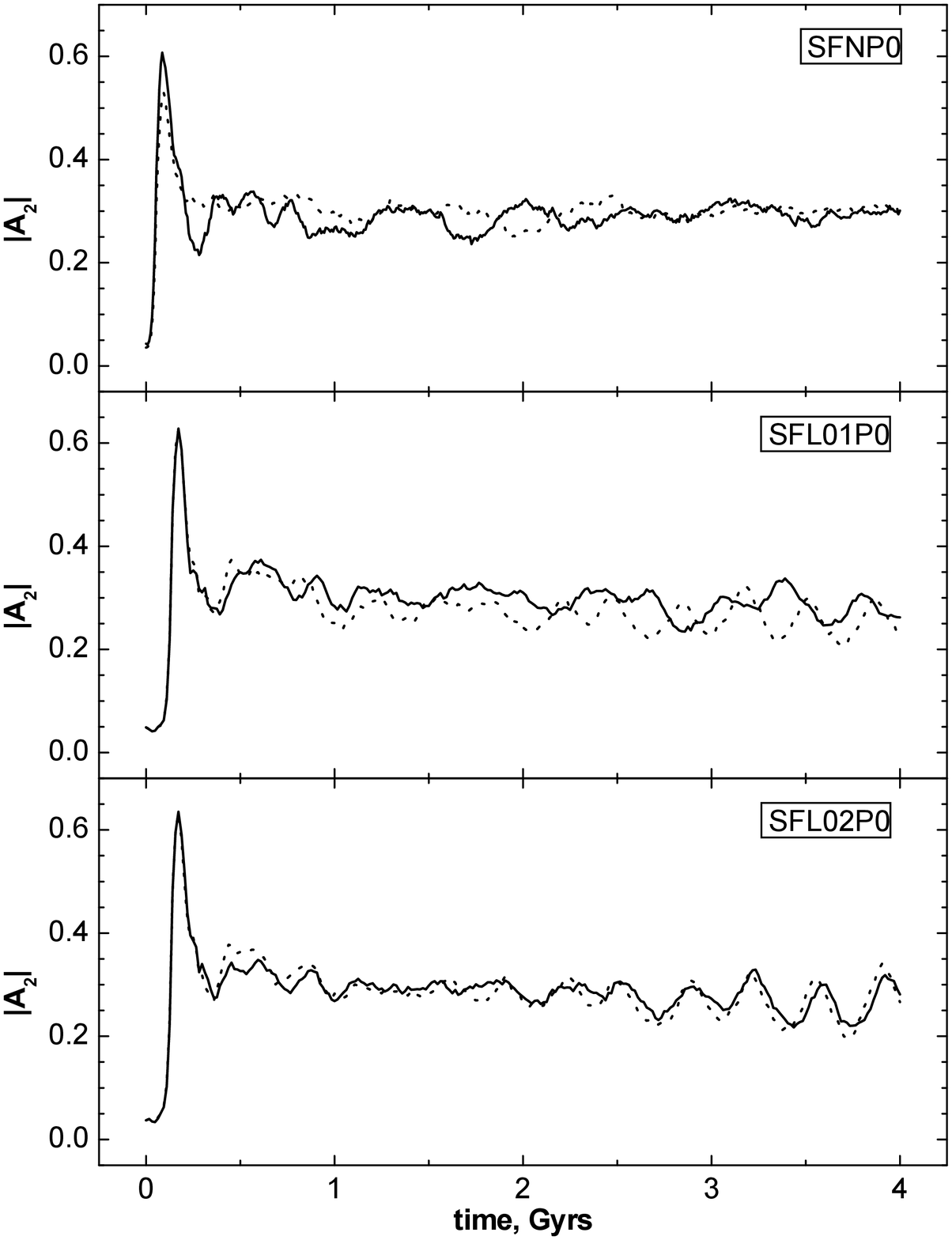}
\includegraphics[width=65mm]{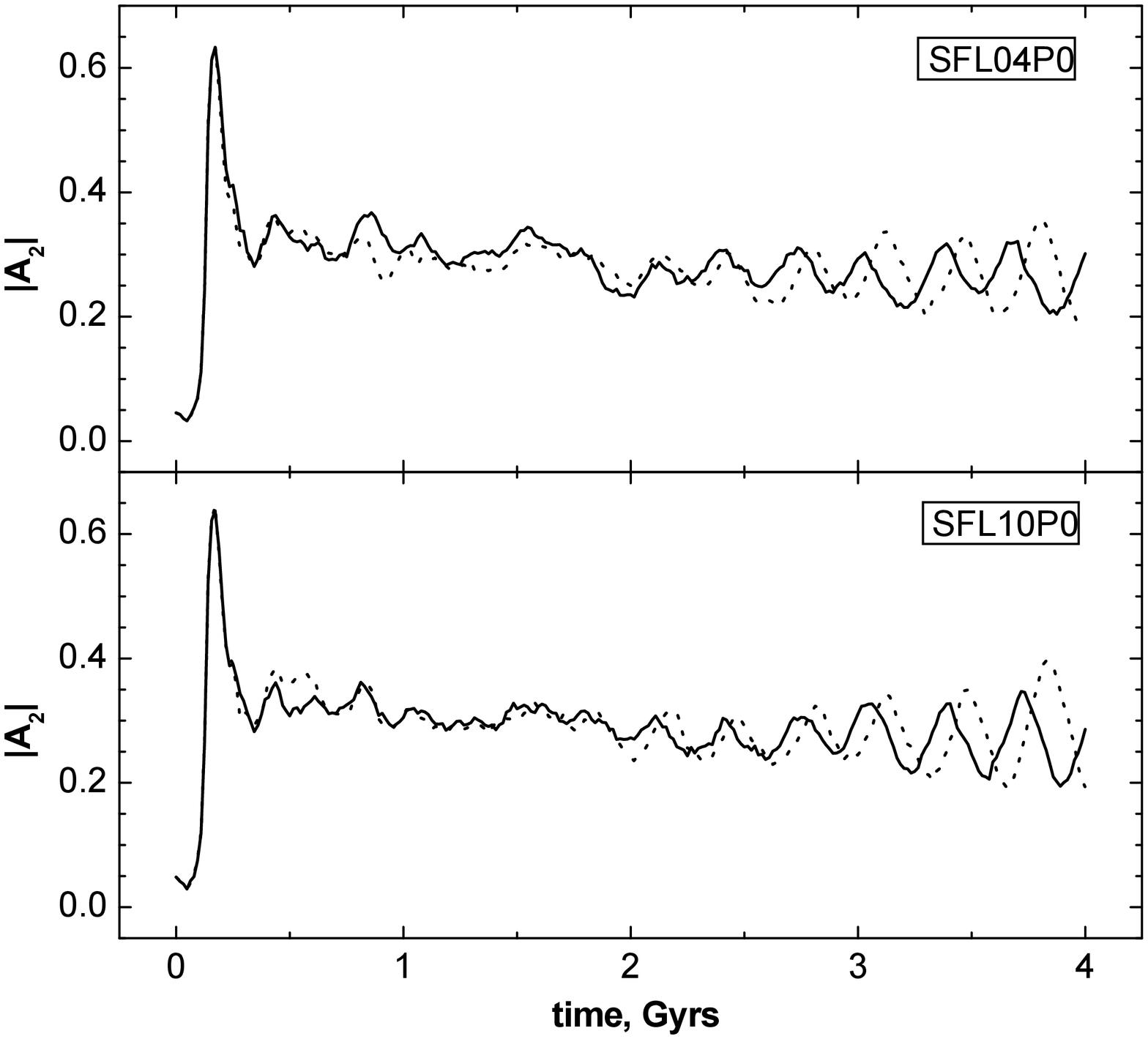}
\vspace{-2.5cm}
\caption{Evolution of $|A_2|$ for models SFNP0-SFL10P0. Dotted
lines correspond to the first galaxy (retrograde orbit) whereas
solid lines correspond to the second galaxy (prograde orbit).}
\label{sf-harms0R}
\end{center}
\end{figure}
%
\begin{figure}
\begin{center}
\includegraphics[width=65mm]{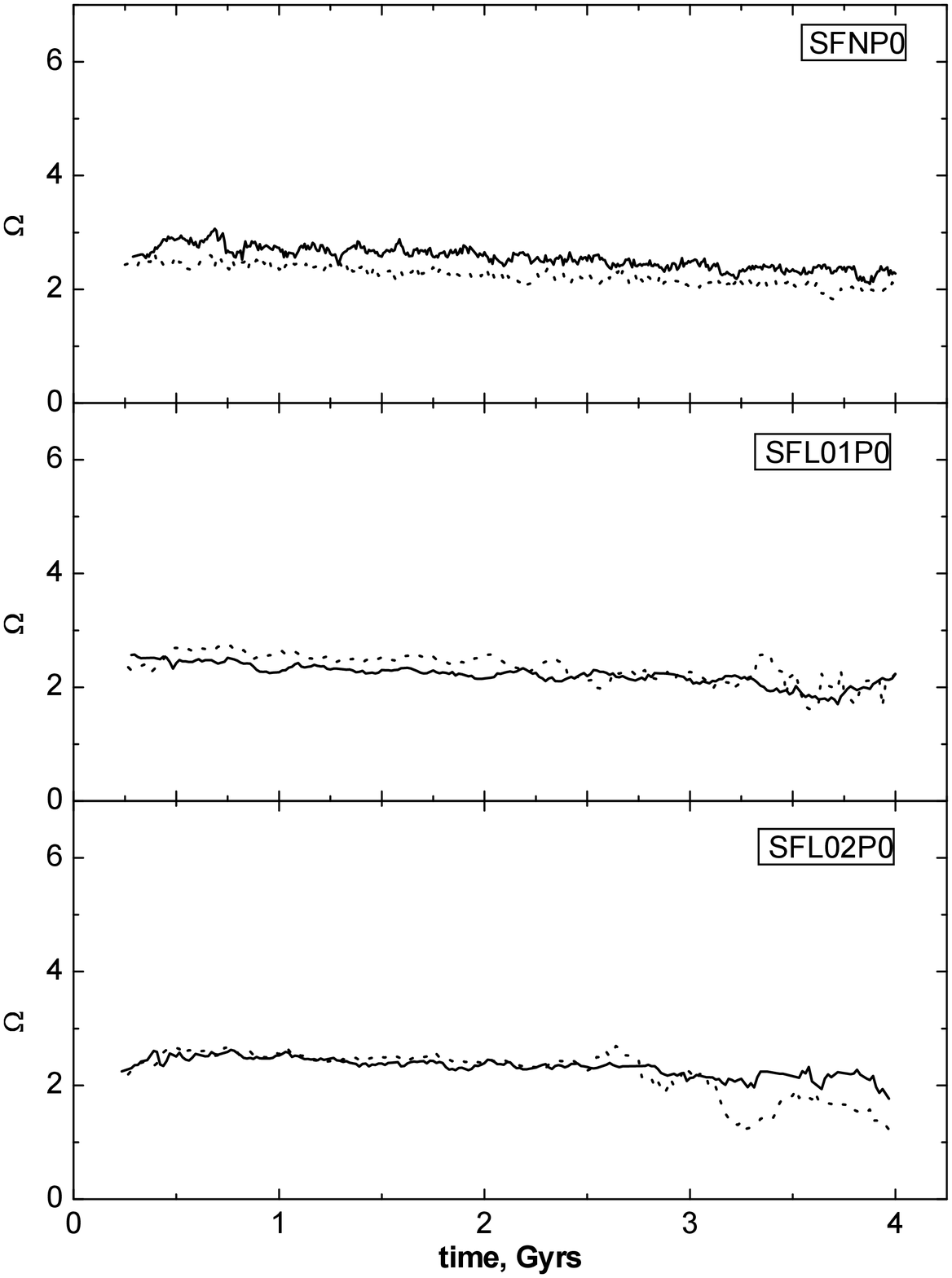}
\includegraphics[width=65mm]{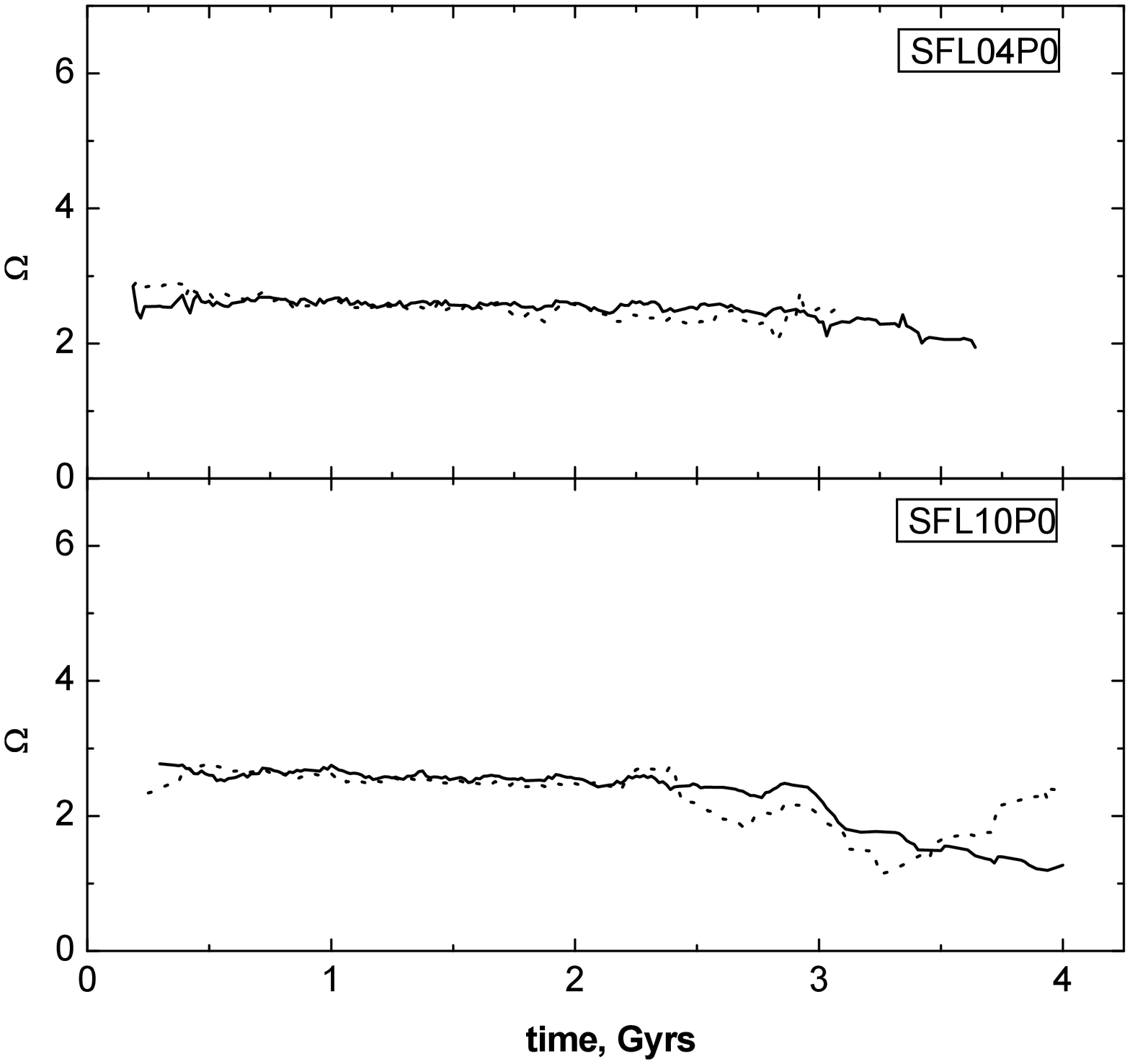}
\vspace{-2.5cm}
\caption{Evolution of $\Omega$ for models SFNP0-SFL10P0. The
correspondence of curves is the same as in Fig.~\ref{sf-harms0R}.}
\label{sf-omegas0R}
\end{center}
\end{figure}
%
\begin{figure}[t]
\begin{center}
\includegraphics[width=65mm]{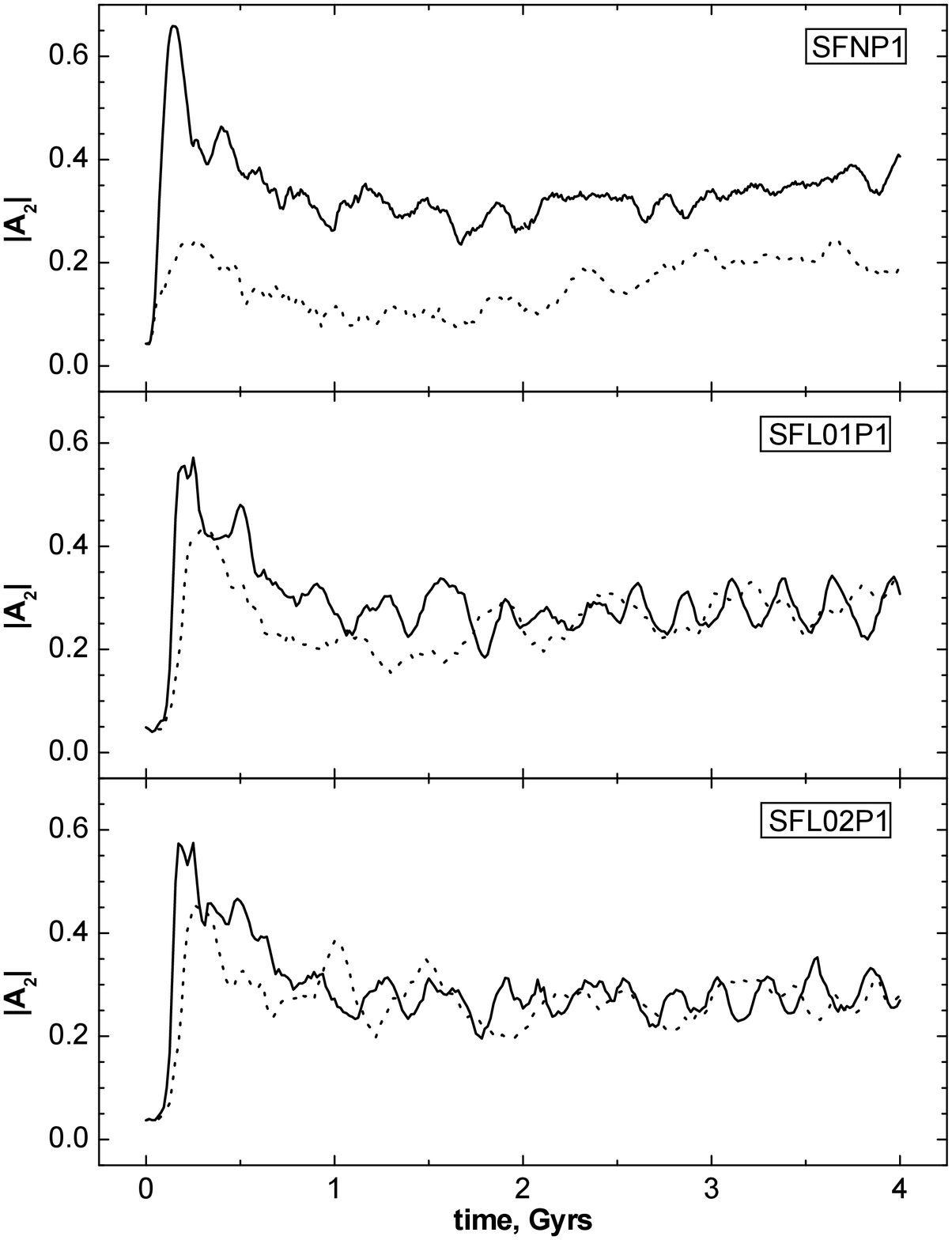}
\includegraphics[width=65mm]{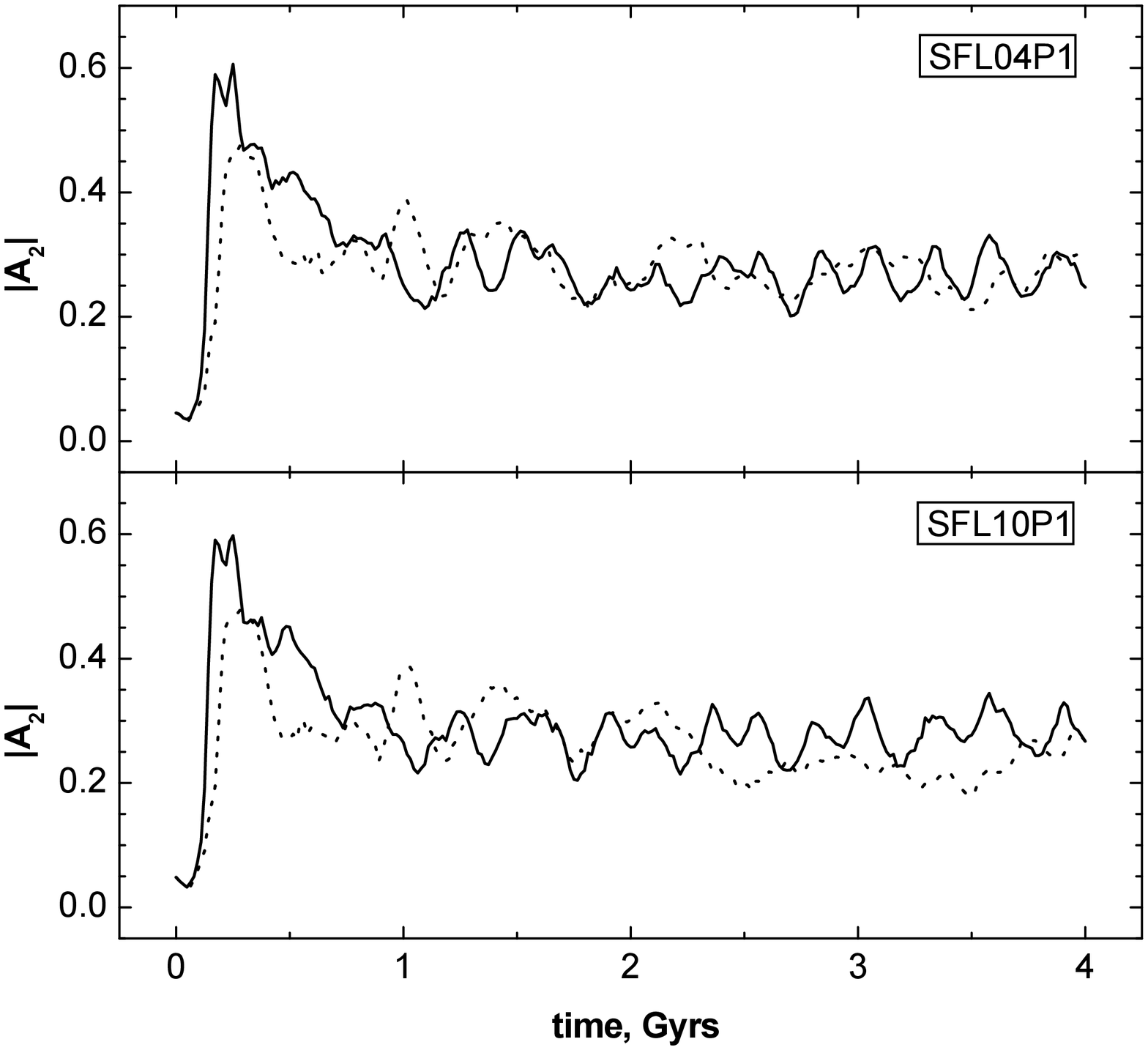}
\vspace{-2.5cm}
\caption{Evolution of $|A_2|$ for models SFNP1-SFL10P1. The
correspondence of curves is the same as in Fig.~\ref{sf-harms0R}.}
\label{sf-harms1R}
\end{center}
\end{figure}
%
\begin{figure}[t]
\begin{center}
\includegraphics[width=65mm]{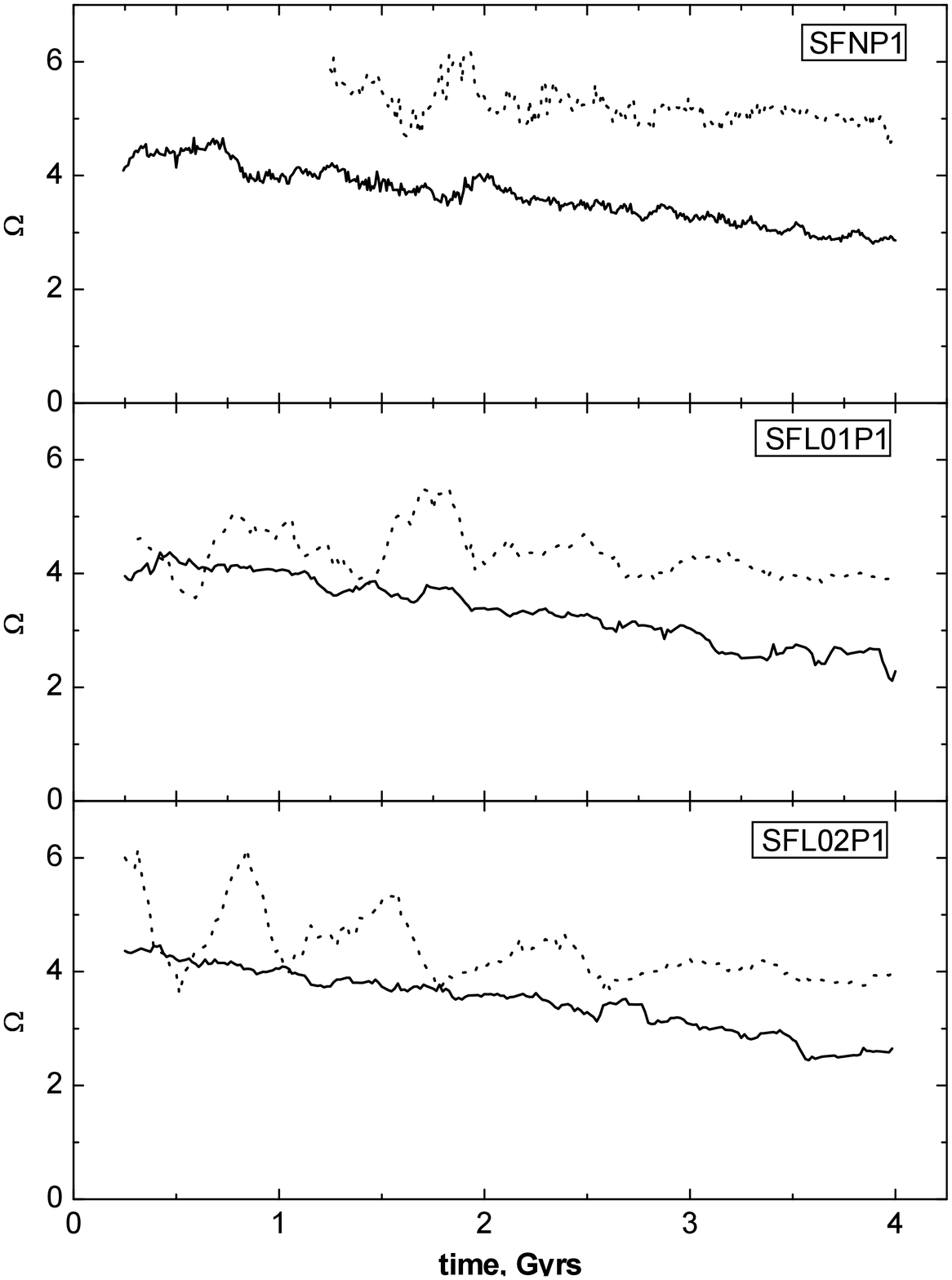}
\includegraphics[width=65mm]{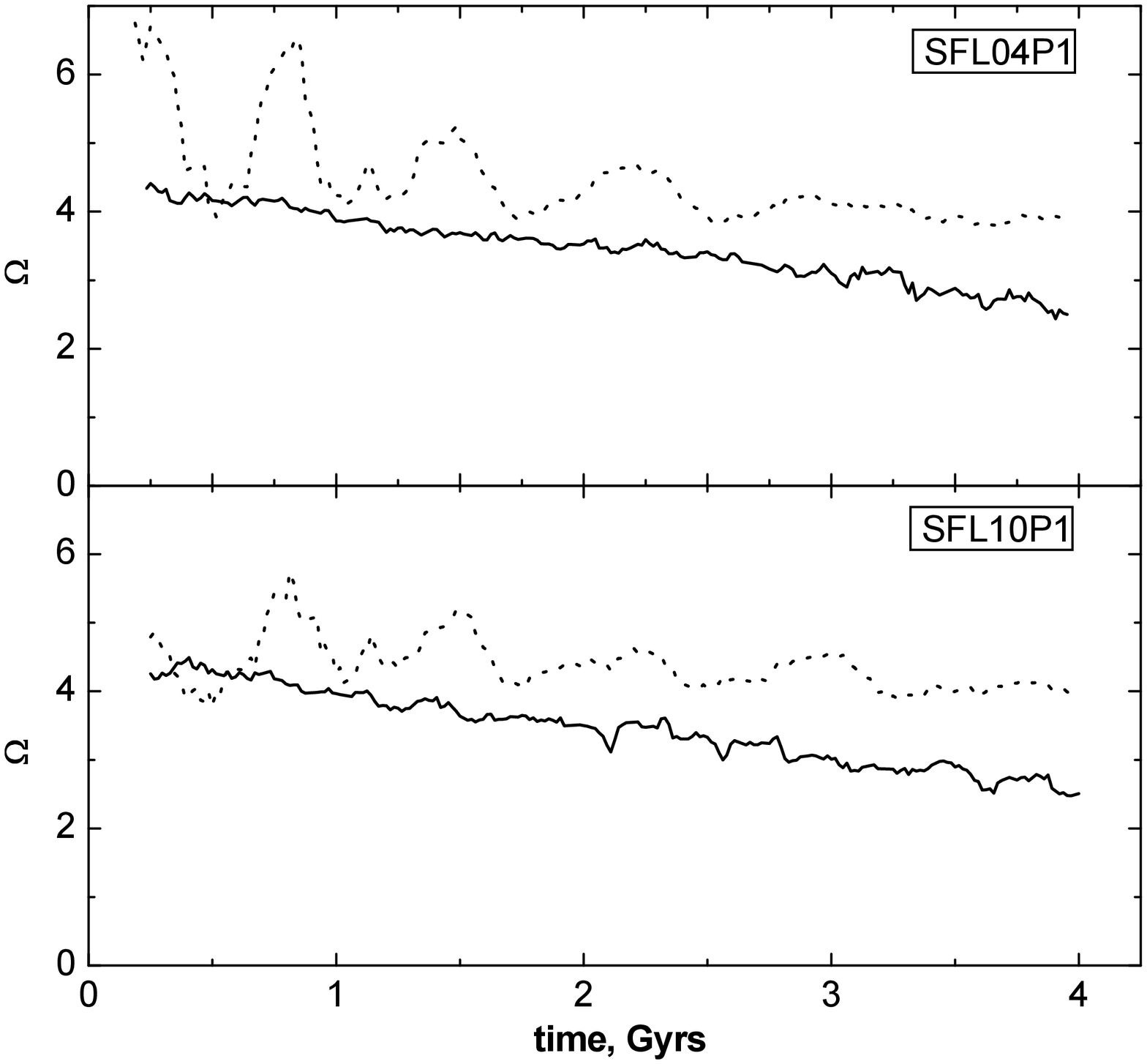}
\vspace{-2.5cm}
\caption{Evolution of $\Omega$ for models SFNP1-SFL10P1. The
correspondence of curves is the same as in Fig.~\ref{sf-harms0R}.}
\label{sf-omegas1R}
\end{center}
\end{figure}
%
\begin{figure}[t]
\begin{center}
\includegraphics[width=65mm]{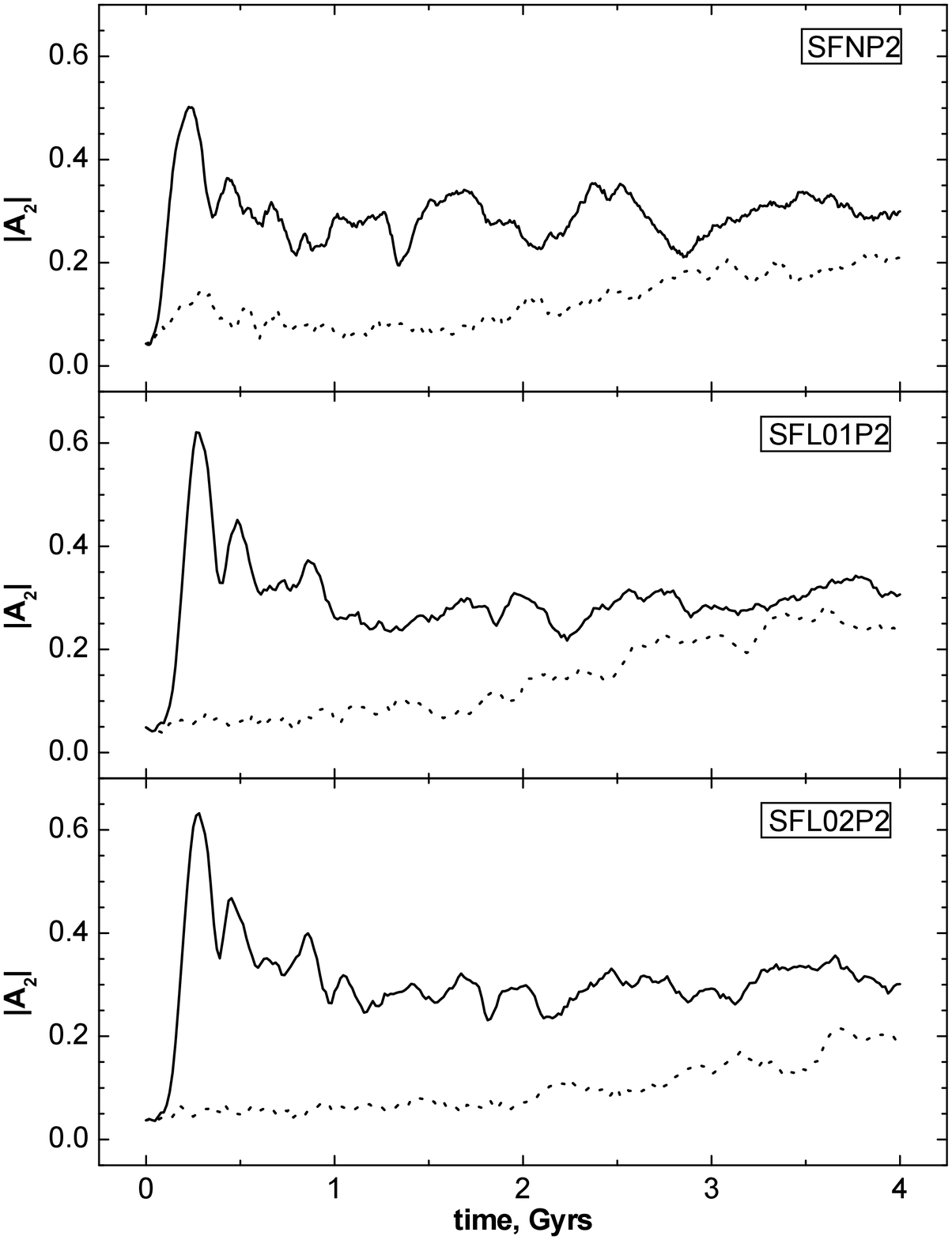}
\includegraphics[width=65mm]{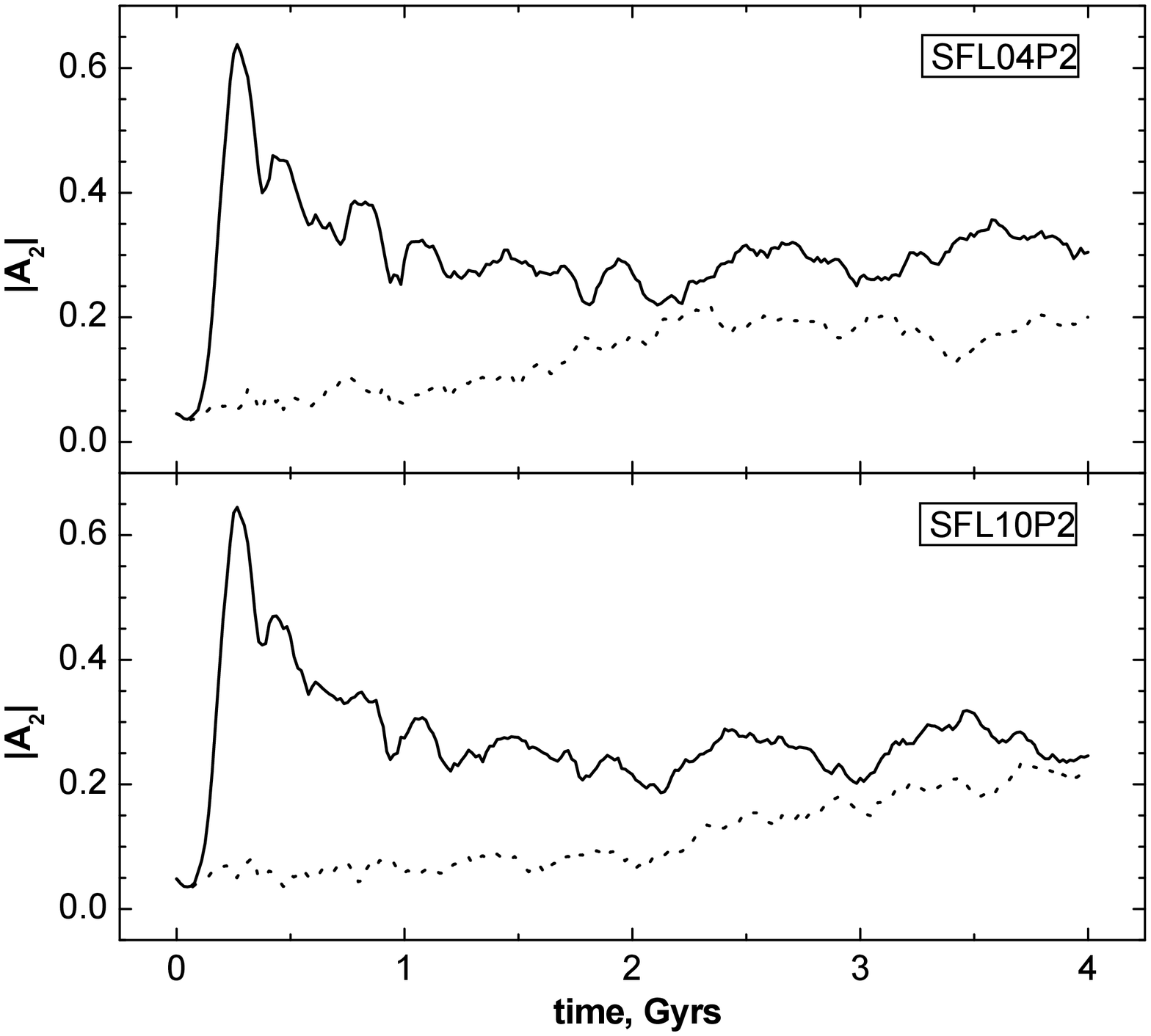}
\vspace{-2.5cm}
\caption{Evolution of $|A_2|$ for models SFNP2-SFL10P2. The
correspondence of curves is the same as in Fig.~\ref{sf-harms0R}.}
\label{sf-harms2R}
\end{center}
\end{figure}
%
\begin{figure}
\begin{center}
\includegraphics[width=65mm]{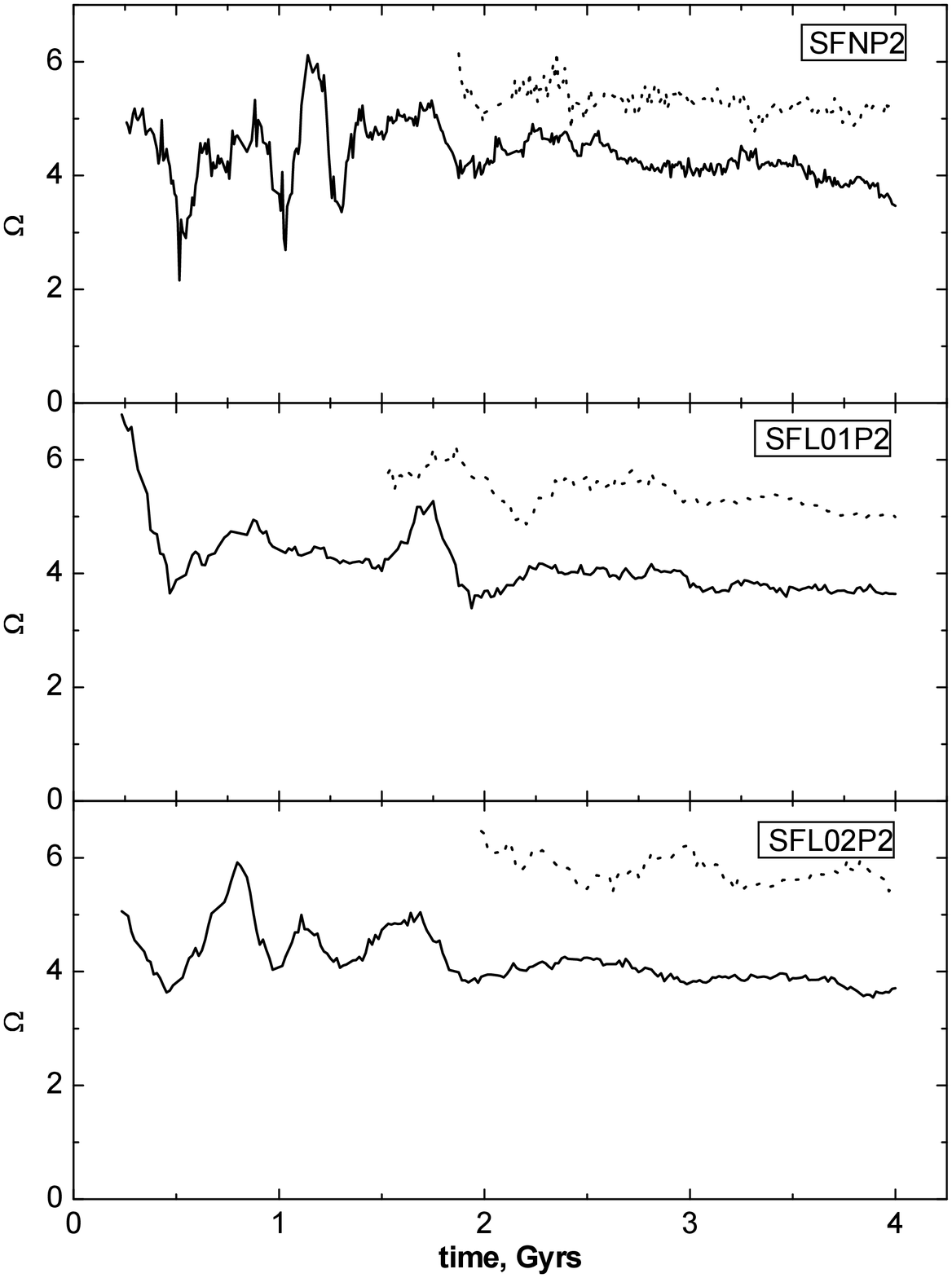}
\includegraphics[width=65mm]{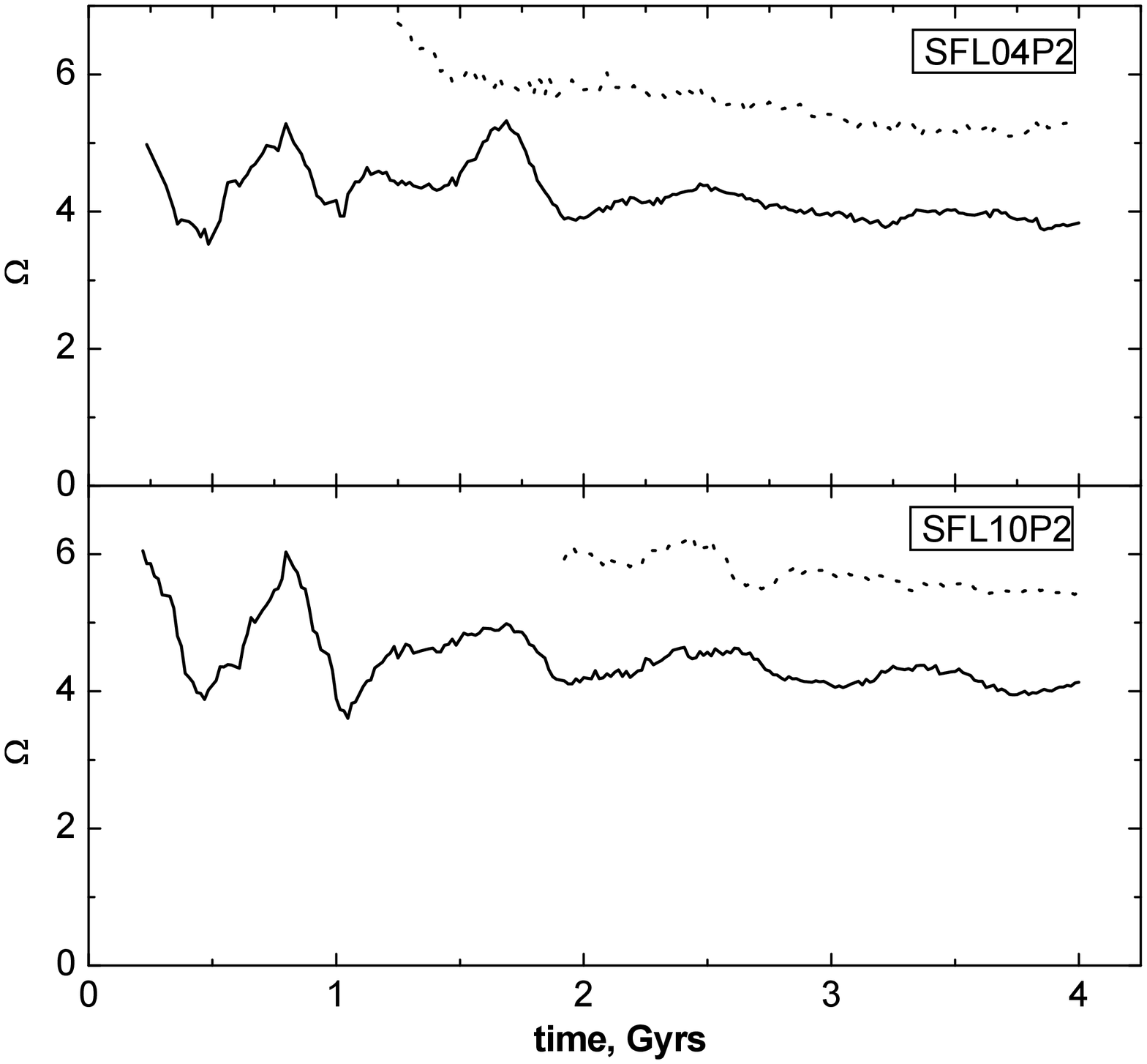}
\vspace{-2.5cm}
\caption{Evolution of $\Omega$ for models SFNP2-SFL10P2. The
correspondence of curves is the same as in Fig.~\ref{sf-harms0R}.}
\label{sf-omegas2R}
\end{center}
\end{figure}
%
%
In general, the presence of SF with positive $\alpha$ reduces the
gravitation force on scales $r>\lambda$. In Ref. \cite{Gabbasov2006a} we 
have seen that larger values of 
$\varepsilon$ for the Plummer softening produce bars earlier. 
Thus, it is possible that the reduction of gravity
either due to small $\varepsilon$ or large $\lambda$ is
responsible for early formation of long bars. With negative
$\alpha$ we expect to have a contrary result, i.e., smaller
bars that form later. 

In addition, we have looked for morphological differences in spiral arms,
such as the {\it wiggle} effect, which is the periodic change of the direction of short arms at the end of
the bar from leading to trailing \cite{Gabbasov2006a}.
We visually search for the presence of the wiggle in both discs, and the results are listed in Table \ref{SF-collisions}.  
\section{Conclusions}\label{conclusions}
We have used the Newtonian limit of general STT that are compatible with local observations by
the appropriate definition of the background field constant, i.e., $<\phi> = G_{N}/(1+\alpha)$.  Then, 
from large-scale experiments we set a range of possible variations of the parameters of the modified 
gravitational theory,  parametrized by $(\lambda, \alpha)$.  The general gravitational effect is that  
the interaction with the SF becomes weaker ($\alpha >0$) by a factor $1/(1+\alpha)$ for $r>\lambda$  in 
comparison with the Newtonian case. 
Using  the resulting modified dynamics,  we have studied isolated spirals and the collision of
two equal spiral galaxies.  From our simulations  with different
$\lambda$, we have found that  the inclusion of the SF changes the dynamical properties
of galaxies such as the bar morphology and pattern velocity.

From the performed simulations of isolated galaxy models with different
$\lambda$, we can see that the addition of a non-minimally coupled
SF slightly modifies the equilibrium of Newtonian model,
 acting as a perturbation, and diminishes the total potential energy,
 since the effective gravitational constant diminishes.  This effect destabilizes 
 the disc to form a bar in all models with $N=163\,840$. For $\alpha=0.1$, the 
 SF interaction scale $\lambda=16$ kpc produces a strongest bar, while
for $\alpha=0.3$, a strong bar forms for $\lambda=8$ kpc. Also,
for these scales bars appear earlier. This suggests that there
exists some kind of resonance between stellar orbits and SF
interaction scale.   The results we have found  with positive values
of $\alpha$ imply that most of the spiral galaxies should be barred, 
but this does not exactly correspond with the observational fact that around  $70\%$ 
of isolated galaxies are barred \cite{ElElHi04}.  However, the bar found 
might be the result of the algorithm  to construct the initial models. 
Therefore,  a next step is  to construct a stable self-consistent 
model in accordance with the modified gravity.   On the other hand, one could also study 
the effects for negative values of $\alpha$, where the force augments for distances 
bigger than $\lambda$.    A wide range of parameters should be investigated and higher resolution have
 to be used in simulations in order to make predictions for particular models.

In the study of isolated galaxies  was shown that  
the presence of the SF destabilizes the disc of isolated galaxies and favors the bar formation. 
For collisions of two galaxies we observe the same trend.
In the off-axis collisions with the impact parameter equal to the disc radius, 
the bars in both prograde and retrograde discs have the same amplitude, 
independently of $\lambda$.  However, the wiggle does not appear in the second disc, 
as shown in Table \ref{SF-collisions}.
All these properties depend on the pair ($\lambda$, $\alpha$), which,
 on the other hand, can be constrained from observations that eventually will discriminate  
 among the different values of the parameters of the theory. 
 
 The results presented are only preliminary, and we  describe the overall differences
 without giving a full interpretation. A broad range of parameters should be investigated and higher resolution have
 to be used in simulations in order to make comparisons with the observed interacting galaxies.
 
 \bigskip
 
 {\it Acknowledgements: } This work has been partially supported by CONACYT  under  
contracts U43534-R, 44917-F and J200.476/2004. 
\vfill 


\appendix*
\section{The Newtonian limit of STT}\label{sec:sfderiv}
We start with the field equations are given bye Eqs.  
(\ref{Einsteineqn}) and (\ref{SFparteqn}) and compute the Newtonian
limit of the scalar-tensor theory to first order. Accordingly, we assume that the
potential oscillates around the background field
\begin{equation}
\phi=\left<\phi\right>+\bar{\phi},\qquad
\left<\phi\right>=\mbox{const}.\,,
\end{equation}
and we expand the field quantities around $\left<\phi\right>$ and
the Minkowski metric ($g_{\mu\nu}=\eta_{\mu\nu}+h_{\mu\nu}$) to
first order, being $\left<\phi\right> << \bar\phi$ and
$h_{\mu\nu}<<\eta_{\mu\nu}$. We assume that the field is
quasi-stationary, such that all time derivatives can be ignored:
$\square\phi=-\nabla^2\bar\phi$.   
Additionally, one has that  $\rho c^2>>p$, and $T=T_{00}=\rho$.

With these assumptions we can expand all terms of field equations
(\ref{Einsteineqn}-\ref{SFparteqn}) and use Taylor series for the
unknown functions $\omega(\phi)$ and $V(\phi)$ in terms of the
small quantity $\bar{\phi}$. Let us first consider
Eq.\ (\ref{Einsteineqn}). The terms on the l.h.s., after inserting the
perturbed metric and assuming the linearized  harmonic gauge, can be written as:
\begin{equation}
\label{lhs-expansion}
R_{\mu\nu}-\frac{1}{2}g_{\mu\nu}R=\frac{1}{2}\left(h_{\mu\nu}-
\frac{1}{2}\eta_{\mu\nu}h\right)^{,\lambda}_{\ ,\lambda},
\end{equation}
where $h=h^{\mu}_{\mu}$. The first term on the r.h.s. is the usual
general relativity term. The second term can be written as
follows:
\begin{equation}
\label{Vg-expansion} \frac{1}{2}Vg_{\mu\nu}=\frac{1}{2}V
\left(\eta_{\mu\nu}+h_{\mu\nu}\right).
\end{equation}
The third and fourth terms vanish because these are second order
terms in the expansion:
\begin{equation}
\partial_{\mu}\phi\partial_{\nu}\phi=\partial_{\mu}\bar{\phi}\partial_{\nu}\bar{\phi}.
\end{equation}
The last two terms can be written as
\begin{equation}
\phi_{;\mu\nu}-g_{\mu\nu}\square\phi=\bar{\phi}_{;\mu\nu}+\eta_{\mu\nu}\nabla^2\bar{\phi},
\end{equation}
and
\begin{equation}
\bar{\phi}_{;\mu\nu}=\left(\bar{\phi}_{,\mu}\right)_{;\nu}=\bar\phi_{,\mu\nu}-
\Gamma^{\beta}_{\mu\nu}\bar\phi_{,\beta}.
\end{equation}
Substituting the above expansions into Eq.\ (\ref{Einsteineqn}) we may
write:
\begin{eqnarray}
\label{fieldeq1-expanded} \frac{1}{2}\left(h_{\mu\nu}-
\frac{1}{2}\eta_{\mu\nu}h\right)^{,\lambda}_{\ ,\lambda}=-
\left[\frac{1}{\left<\phi\right>}-\frac{1}{\left<\phi\right>^2}\bar{\phi}\right]\times
\nonumber \\
\left[8\pi T_{\mu\nu}-\frac{V}{2}\left(\eta_{\mu\nu}+
h_{\mu\nu}\right)+\bar\phi_{;\mu\nu}+\left(\eta_{\mu\nu}+
h_{\mu\nu}\right)\nabla^2\bar{\phi}\right] \;\;\;
\end{eqnarray}
If we multiply the Eq.~(\ref{fieldeq1-expanded}) by
$\eta^{\mu\nu}$ and take the trace, the l.h.s. can be written as
\begin{equation}
\frac{1}{2}\left(\eta^{\mu\nu}h_{\mu\nu}-
\frac{1}{2}\eta^{\mu\nu}\eta_{\mu\nu}h\right)^{,\lambda}_{\
,\lambda}= \frac{1}{2}\left(h-2h\right)^{,\lambda}_{,\lambda}=
-\frac{1}{2}h^{,\lambda}_{\ ,\lambda}\,,
\end{equation}
and the r.h.s. as
\begin{eqnarray}
\label{fieldeq1-rhs-expanded}
-\left[\frac{1}{\left<\phi\right>}-\frac{1}{\left<\phi\right>^2}\bar{\phi}\right] 
\times \nonumber \\
\left[8\pi\rho-\frac{V}{2}(4+h)-\nabla^2\bar\phi+(4+h)\nabla^2\bar\phi\right].
\end{eqnarray}
In the last equation we used the relation
\begin{equation}
\bar\phi^{\quad ;\nu}_{,\nu}=-\nabla^2\bar\phi,
\end{equation}
valid to first order.
 Now, let us rewrite the
Eq.~(\ref{lhs-expansion}) as follows:
\begin{equation}
\label{lhs-derivative} \frac{1}{2}\left(h_{\mu\nu}-
\frac{1}{2}\eta_{\mu\nu}h\right)^{,\lambda}_{\ ,\lambda}=
\frac{1}{2}\left(h^{,\lambda}_{\mu\nu,\lambda}-
\frac{1}{2}\eta_{\mu\nu}h^{,\lambda}_{\ ,\lambda}\right),
\end{equation}
in which we insert \ref{fieldeq1-rhs-expanded} instead of the term
$-h^{,\lambda}_{\ ,\lambda}$ and using
Eq.~(\ref{fieldeq1-expanded}) we obtain
\begin{eqnarray}
\label{fieldeq1-rhs-expandeddd}
\frac{1}{2}h^{,\lambda}_{\mu\nu,\lambda}-\frac{1}{2}\eta_{\mu\nu}
\left[\frac{1}{\left<\phi\right>}-\frac{1}{\left<\phi\right>^2}\bar{\phi}\right]
\times 
\nonumber \\
\left[8\pi\rho-\frac{V}{2}(4+h)+h\nabla^2\bar\phi+3\nabla^2\bar\phi\right]=
\nonumber \\
-\left[\frac{1}{\left<\phi\right>}-\frac{1}{\left<\phi\right>^2}\bar{\phi}\right]
\times 
\nonumber \\
\left[8\pi
T_{\mu\nu}-\frac{V}{2}\left(\eta_{\mu\nu}+h_{\mu\nu}\right)
+\bar\phi_{;\mu\nu}+\left(\eta_{\mu\nu}+h_{\mu\nu}\right)\nabla^2\bar\phi\right].
\nonumber \\ 
\end{eqnarray}
Let $\mu=\nu=0$. Then, the above equation becomes to first order
\begin{eqnarray}
\label{fieldeq1-00} \frac{1}{2}h^{,\lambda}_{00,\lambda}\equiv
-\nabla^2\Phi=
\nonumber \\
- \left[\frac{1}{\left<\phi\right>}\right]
\left[4\pi\rho+\frac{V}{2}-\frac{V}{2}h_{00}+\frac{V
h}{4}-\frac{1}{2}\nabla^2\bar\phi\right]{}
\nonumber \\
+\left[\frac{\bar{\phi}}{\left<\phi\right>^2}\right] \left[4\pi
\rho+\frac{V}{2}\right],
\end{eqnarray}
where $\Phi$ is modified Newtonian potential. Here we used the
fact that
\begin{eqnarray}
\frac{1}{2}\left(h_{00}\right)^{,\lambda}_{\ ,\lambda}=
\frac{1}{2}\left(h_{00}\right)^{,0}_{\
,0}-\frac{1}{2}\left(h_{00}\right)^{,i}_{\ ,i}=
-\frac{1}{2}\left(h_{00}\right)^{,i}_{\ ,i}=
\nonumber \\
\frac{1}{2}\nabla^2
h_{00} .\qquad
\end{eqnarray}
Finally, expanding the potential
\begin{equation}
\label{V-expand}
V=V(\left<\phi\right>)+ \left.\frac{\partial
V}{\partial
\bar{\phi}}\right|_{\langle\phi\rangle}\bar{\phi}+\cdots,
\end{equation}
and discarding terms higher than first order we reach the the
expression
\begin{equation}
\label{fieldeq1-limit} \frac{1}{2}\nabla^2 \Phi_N =
\frac{1}{\langle\phi\rangle} \left[ 4\pi \rho - \frac{1}{2}
\nabla^2 \bar{\phi}\right]  \; .
\end{equation}
 Accordingly, particles in our galactic
models experience a force given by ${\bf F}=-\nabla\Phi_N$.
Now, consider Eq.~(\ref{SFparteqn}). The terms that depend on
$\phi$ can be expanded to the first order:
\begin{equation}
\label{a-expansion}
\frac{1}{3+2\omega}=\left.\frac{1}{3+2\omega}\right|_{(\langle\phi\rangle)}
+\left.\frac{\partial}{\partial\bar{\phi}}\left(\frac{1}{3+2\omega}\right)
\right|_{\langle\phi\rangle}\bar{\phi}+\cdots
\end{equation}
\begin{equation}
\label{omega-expansion}
\omega^{'}(\phi)=\omega^{'}_{\langle\phi\rangle}+\omega^{''}_{\langle\phi\rangle}
\bar{\phi}+\cdots
\end{equation}
Let us make the following notation:
\begin{equation}
V(\left<\phi\right>)\equiv V_{\left<\phi\right>}, \quad
\left.\frac{\partial V}{\partial
\bar{\phi}}\right|_{\langle\phi\rangle}\equiv
V^{'}_{\left<\phi\right>}.
\end{equation}
\begin{equation}
\label{notation2}
\left.\frac{1}{3+2\omega}\right|_{(\langle\phi\rangle)}\equiv
\alpha_{\langle\phi\rangle}, \quad
\left.\frac{\partial}{\partial\bar{\phi}}\left(\frac{1}{3+2\omega}\right)
\right|_{\langle\phi\rangle}\equiv
\alpha^{'}_{\langle\phi\rangle}.
\end{equation}
Substituting Eqs.~(\ref{V-expand}),
(\ref{a-expansion}-\ref{notation2}) into Eq.~(\ref{SFparteqn}) and
keeping only the terms up to the first order, we obtain:
\begin{eqnarray}
\label{fieldeq2-reduced}
-\nabla^2\bar{\phi}
+\alpha^2_{\langle\phi\rangle}
\left[
\alpha^{-1}_{\langle\phi\rangle}
\left({\left<\phi\right>}V^{''}_{\left<\phi\right>}-V^{'}_{\left<\phi\right>}\right)
\right. \nonumber \\ \left.
-2\omega^{'}_{\langle\phi\rangle}\left({\langle\phi\rangle}V^{'}_{\left<\phi\right>}-
2V_{\left<\phi\right>}\right)
+16\pi\rho\omega^{'}_{\langle\phi\rangle}
\right]
\bar{\phi} \nonumber \\
{}=\alpha_{\langle\phi\rangle}
\left[8\pi\rho+2V_{\langle\phi\rangle}-{\langle\phi\rangle}
V^{'}_{\left<\phi\right>}\right].
\end{eqnarray}
The constant in the second term of the l.h.s. represents an
-squared- effective mass term of the theory that we will denote as
$m^2$. Last two terms in the r.h.s. represent a cosmological
constant that we will set to zero, since the mean density of the
galaxy is much greater than this term. Accordingly, we finally
have:
\begin{equation} 
\label{fieldeq2-limit} -\nabla^2\bar{\phi}+m^2\bar{\phi}=
8\pi\alpha_{\langle\phi\rangle} \rho \, .
\end{equation}
This equation together with
Eq.~(\ref{fieldeq1-limit}) represent the Newtonian limit of
general STT that can be expanded around
the background quantities $\langle\phi\rangle$ and
$\eta_{\mu\nu}$.  

\end{document}